\magnification=\magstep1
\input amssym.tex
\font\bigbf=cmbx10 scaled\magstep1

 at 10truept


\def\frac#1#2{{#1\over #2}}

\def\proc #1. #2\par{\medbreak
\noindent{\bf#1. \enspace}{\rm#2}\par\medbreak}

{\par\medskip}
\footline={\ifnum\pageno=1{\hfil}\else{\hss\tenrm\folio\hss}\fi}
\mathsurround=2pt
\tolerance=20000

\newread\epsffilein    
\newif\ifepsffileok    
\newif\ifepsfbbfound   
\newif\ifepsfverbose   
\newif\ifepsfdraft     
\newdimen\epsfxsize    
\newdimen\epsfysize    
\newdimen\epsftsize    
\newdimen\epsfrsize    
\newdimen\epsftmp      
\newdimen\pspoints     
\pspoints=1bp          
\epsfxsize=0pt         
\epsfysize=0pt         
\def\epsfbox#1{\global\def\epsfllx{72}\global\def\epsflly{72}%
   \global\def\epsfurx{540}\global\def\epsfury{720}%
   \def\lbracket{[}\def\testit{#1}\ifx\testit\lbracket
   \let\next=\epsfgetlitbb\else\let\next=\epsfnormal\fi\next{#1}}%
\def\epsfgetlitbb#1#2 #3 #4 #5]#6{\epsfgrab #2 #3 #4 #5 .\\%
   \epsfsetgraph{#6}}%
\def\epsfnormal#1{\epsfgetbb{#1}\epsfsetgraph{#1}}%
\def\epsfgetbb#1{%
%
%
\openin\epsffilein=#1
\ifeof\epsffilein\errmessage{I couldn't open #1, will ignore it}\else
%
%
   {\epsffileoktrue \chardef\other=12
    \def\do##1{\catcode`##1=\other}\dospecials \catcode`\ =10
    \loop
       \read\epsffilein to \epsffileline
       \ifeof\epsffilein\epsffileokfalse\else
%
%
          \expandafter\epsfaux\epsffileline:. \\%
       \fi
   \ifepsffileok\repeat
   \ifepsfbbfound\else
    \ifepsfverbose\message{No bounding box comment in #1; using defaults}\fi\fi
   }\closein\epsffilein\fi}%
%
%
%
\def\epsfclipoff{\def\epsfclipstring{\ifepsfdraft\space clip\fi}}%
\epsfclipoff
\def\epsfsetgraph#1{%
   \epsfrsize=\epsfury\pspoints
   \advance\epsfrsize by-\epsflly\pspoints
   \epsftsize=\epsfurx\pspoints
   \advance\epsftsize by-\epsfllx\pspoints
%
%
   \epsfxsize\epsfsize\epsftsize\epsfrsize
   \ifnum\epsfxsize=0 \ifnum\epsfysize=0
      \epsfxsize=\epsftsize \epsfysize=\epsfrsize
      \epsfrsize=0pt
%
%
     \else\epsftmp=\epsftsize \divide\epsftmp\epsfrsize
       \epsfxsize=\epsfysize \multiply\epsfxsize\epsftmp
       \multiply\epsftmp\epsfrsize \advance\epsftsize-\epsftmp
       \epsftmp=\epsfysize
       \loop \advance\epsftsize\epsftsize \divide\epsftmp 2
       \ifnum\epsftmp>0
          \ifnum\epsftsize<\epsfrsize\else
             \advance\epsftsize-\epsfrsize \advance\epsfxsize\epsftmp \fi
       \repeat
       \epsfrsize=0pt
     \fi
   \else \ifnum\epsfysize=0
     \epsftmp=\epsfrsize \divide\epsftmp\epsftsize
     \epsfysize=\epsfxsize \multiply\epsfysize\epsftmp   
     \multiply\epsftmp\epsftsize \advance\epsfrsize-\epsftmp
     \epsftmp=\epsfxsize
     \loop \advance\epsfrsize\epsfrsize \divide\epsftmp 2
     \ifnum\epsftmp>0
        \ifnum\epsfrsize<\epsftsize\else
           \advance\epsfrsize-\epsftsize \advance\epsfysize\epsftmp \fi
     \repeat
     \epsfrsize=0pt
    \else
     \epsfrsize=\epsfysize
    \fi
   \fi
%
%
   \ifepsfverbose\message{#1: width=\the\epsfxsize, height=\the\epsfysize}\fi
   \epsftmp=10\epsfxsize \divide\epsftmp\pspoints
   \vbox to\epsfysize{\vfil\hbox to\epsfxsize{%
      \ifnum\epsfrsize=0\relax
        \includegraphics{\ifepsfdraft}%
      \else
        \epsfrsize=10\epsfysize \divide\epsfrsize\pspoints
        \includegraphics{\ifepsfdraft}%
      \fi
      \hfil}}%
\global\epsfxsize=0pt\global\epsfysize=0pt}%
%
%
{\catcode`\%=12 \global\let\epsfpercent=
%
%
\long\def\epsfaux#1#2:#3\\{\ifx#1\epsfpercent
   \def\testit{#2}\ifx\testit\epsfbblit
      \epsfgrab #3 . . . \\%
      \epsffileokfalse
      \global\epsfbbfoundtrue
   \fi\else\ifx#1\par\else\epsffileokfalse\fi\fi}%
%
%
\def\epsfempty{}%
\def\epsfgrab #1 #2 #3 #4 #5\\{%
\global\def\epsfllx{#1}\ifx\epsfllx\epsfempty
      \epsfgrab #2 #3 #4 #5 .\\\else
   \global\def\epsflly{#2}%
   \global\def\epsfurx{#3}\global\def\epsfury{#4}\fi}%
%
%
\def\epsfsize#1#2{\epsfxsize}
%
%


\centerline {\bigbf On a neutrino theory of matter}
\vglue .15truein
\vglue .25truein
\centerline{\bf D. K. Sen}
\smallskip

\centerline{Department of Mathematics}
\centerline{University of Toronto}
\centerline{Toronto, ON M5S 2E4, Canada}
\centerline{email: sen@math.toronto.edu}
\bigskip
\medskip

\medskip
\medskip

{\narrower \noindent
Physically observable particles are assumed to result
from an interaction between massless positively and
negatively oriented 2-component Weyl neutrinos. 
A simple quantum mechanical
analysis of a composite  system of Weyl neutrinos of opposite
orientations with a certain specific
interaction shows that such a model can exhibit
 a 2-fold branching and defect in the
total energy of the system, which could then be interpreted as
formation of massive particles. \smallskip}
\bigskip
\bigskip

\noindent {\bf 1.Introduction}

\medskip
\noindent
The standard model has  had considerable success as a unified
theory of all elementary  particles. Together with Higgs
mechanism it is able to explain the existence and masses of
several new bosons. 
\medskip
\noindent
Nevertheless, it cannot be considered as a complete and fully
satisfactory theory of all elementary particles. 
It cannot, for example, explain intrinsicallt 
why the proton is about 1836 times heavier than the electron.
\medskip
\noindent
In the standard model the phenomenon of neutrino oscillation [1]
requires that neutrinos have non-zero mass. 
\medskip
\noindent In 1957  Heisenberg 
[2],[3] tried to formulate (without much
 success) a unified theory of all elementary particles  starting
 from a non-linear 4-component spinor equation with a built in
fundamental constant.
\medskip
\noindent In this paper we [4],[5] suggest that massless 2-component
Weyl neutrinos, instead of 4-component spinors, 
are probably more fundamental than previously thought.
\noindent We consider  
a composite system consisting of a massless positively oriented
2-component Weyl neutrino and a massless negatively oriented
2-component Weyl neutrino 
with a certain specific symmetry-breaking interaction between the two.
\medskip
\noindent
We assume that the observale physical particles manifest
as energy states of the resuting 4-component system.
A simple quantum mechanical treatment shows that
such a model should exhibit 2-fold branching and energy defects,
which could then be interpreted as formation of particles 
of non-zero rest mass.   
\medskip
\noindent
Such a model can also provide a qualitative,alternative non-standard
explanation of the different flavors of a massless  
4-component neutrino and thus of neutrino oscillation without
assuming a neutrino mass.
\medskip
\noindent
The next step would be to consider such a model in the
framework of quantum field theory.
\medskip
\noindent
{\bf 2.Positively and negatively oriented 
 $2$-component Weyl neutrinos}
\medskip
\medskip
\noindent The Weyl equation

$$
(\sigma_k \partial_k + i \partial_4) \varphi = 0 \eqno(1)
$$

\medskip
\noindent describes a massless positively oriented (i.e.
left-handed) $2$-
component Weyl neutrino $\nu_L$.Here $\varphi=
\pmatrix{\varphi_1 \cr \varphi_2}$.
We use space-time coordinates $x_\mu (= x,y,z,
ict), \mu = 1,2,3,4$ and $k = 1,2,3$. The Pauli spin matrices

$$
 \sigma_1=\pmatrix{0&1 \cr 1&0},\sigma_2=\pmatrix{0&-i \cr i&0        }
 ,\sigma_3=\pmatrix{1&0 \cr 0&-1} \eqno(2)
$$

satisfy

$$
\sigma_1\sigma_2=i\sigma_3, \quad \sigma_2\sigma_3=i\sigma_1,         \quad \sigma_3\sigma_1=i\sigma_2, \eqno(3)
$$

Consider now the following equation

$$
(\varrho_k \partial_k + i\partial_4) \chi = 0 \eqno(4)
$$

where the matrices ${\varrho_k}$
$$
 \varrho_1=\pmatrix{1&0 \cr 0&-1},\varrho_2=\pmatrix{0&-i \cr
i&0},\varrho_3=\pmatrix{0&1 \cr 1&0} \eqno(5)
$$

differ from the Pauli spin matrices  ${\sigma_k}$
only in the interchange of the indices $1$ and $3$.
The set ${\varrho_k}$ satisfy

$$
\varrho_1\varrho_2=-i\varrho_3, \quad \varrho_2\varrho_3=-i\varrho_1, \quad \varrho_3\varrho_1=-i\varrho_2, \eqno(6)
$$

Therefore, Eq.(4) describes a massless negatively
oriented (i.e. right-handed) $2$-component Weyl neutrino $\nu_R$.

\noindent {\bf 3.Composite $\nu_L\!{-}\!\nu_R$ system.} 
\medskip

\noindent
Consider first a composite  $\nu_L\!{-}\!\nu_R$ system
without interaction.
The Hamiltonian of a posively oriented (massless) Weyl neutrino $\nu_L$

described by (1) is given by $H_L=-ic\hbar(\sigma\cdot\nabla)$
with $\sigma=(\sigma_1,\sigma_2,\sigma_3)$
and its eigen-functions $\varphi_{E_L}$ satisfy
$$
\left. \eqalign{
H_L\varphi_{E_L}& = E_+\varphi_{E_L} \cr
or \quad (\sigma\cdot\nabla) \varphi_{E_L}& = (iE_L/c\hbar) \varphi_{E_L}}\right\} \eqno(7)
$$

From now on we shall adopt the conventional units in which
$c=\hbar=1$.
\medskip

\noindent
The solutions of (7) are well-known:
$$
\left. \eqalign{
\varphi_{E_L}({\bf x,p})& = a({\bf p})e^{i{\bf x} \cdot {\bf p}} \cr
where \quad {\bf p}^2 \equiv p^2_1+p^2_2+p^2_3&=E^2_L \quad and \quad a({\bf p})=\pmatrix{
p_1-ip_2 \cr E_L-p_3}}\right\}
\eqno(8)
$$
They describe a positively oriented Weyl neutrino $\nu_L$ with energy $E_L$.
\medskip

\noindent The spectrum of $H_L$ is not discrete and hence
the eigenfunctions have to be normalized by the delta-function.
$$
<\varphi_{E_L}({\bf x,p})|\varphi_{E'_+}({\bf x,p'})>=
a({\bf p})^\dagger a({\bf p'})\delta ({\bf p - p'})
\eqno(9)
$$
\medskip

\medskip
\noindent The eigenfunctions are also $\infty$-fold
degenerate, since ${\bf p}$ can take any value on the 
energy shell.We shall remove this degeneracy by integrating
$\varphi_{E_L}({\bf x,p})$ over the energy shell $S^2_{E_L}:
p^2_1+p^2_2+p^2_3=E^2_L$ which is a 2-sphere of radius $E_L$.
We get (with a slight abuse of notation)
$$
\varphi_{E_L}({\bf x})=\int_{S^2_{E_L}} a({\bf p})e^{i{\bf x} \cdot {\bf p}}dS_p \eqno(10)
$$
as a surface integral over $S^2_{E_L}$.Furthermore
we shall suppose that $\varphi_{E_L}({\bf x})$ are normalised.
\medskip

\noindent Similarly, the corresponding eigenfunctions for the
negatively oriented Weyl neutrino $\nu_R$  with Hamiltonian $H_R=-i(\rho\cdot\nabla)$ with energy $E_R$ are given by
$$
\left . \eqalign{
\chi_{E_R}({\bf y},{\bf q})& = b({\bf q})e^{i{\bf y} \cdot {\bf q}} \cr
where \quad q^2_1+q^2_2+q^2_3& =E^2_R \quad and \quad b({\bf q})=\pmatrix{
q_3-iq_2 \cr E_R-q_1}}\right\}
\eqno(11)
$$
We use $\bf y$ for the coordinate of $\nu_R$ and note that 
$b({\bf q})$ differs from $a({\bf p})$ by an interchange of
the indices $1$ and $3$.A similar integration over the energy
shell $S^2_{E_R}:q^2_1+q^2_2+q^2_3 =E^2_R$ gives
$$
\chi_{E_R}({\bf y}) = \int_{S^2_{E_R}}b({\bf q})e^{i{\bf y} \cdot {\bf q}} dS_q \eqno(12)
$$

\noindent Let $\cal H_L$ and $\cal H_R$ be the respective
Hilbert spaces for $\nu_L$ and $\nu_R$. Consider now a
composite $\nu_L\!{-}\!\nu_R$ system without interaction given 
by the tensor product $\cal H = \cal H_L \otimes \cal H_R$ with the 
Hamiltonian $H_o = H_L + H_R$ (or more precisely
$H_L\otimes I + I\otimes H_R$).
Since $H_o = -i(\sigma \cdot \nabla_+) -i(\rho \cdot \nabla_-)$, where $\nabla_+,\nabla_-$ act on ${\bf x,y}$ respectively, and thus does not depend on $\bf x$ and $\bf y$ explicitly,
not only
$$
\Psi_1 ({\bf x,y}) = \varphi_{E_L}({\bf x}) \otimes \chi_{E_R}({\bf y}) \eqno(13)
$$
but also
$$
\Psi_2 ({\bf x,y}) = \varphi_{E_R}({\bf x}) \otimes \chi_{E_L}({\bf y}) \eqno(14)
$$
are both eigenfunctions of $H_o$ with energy $E_L + E_R$.
Thus the composite system $\nu_L\!{-}\!\nu_R$ behaves like a system
of identical particles even though $\nu_l$ is not identical
to $\nu_R$.

\medskip

\noindent
 {\it We remark that if the universe were spatially
non-orientable, $\nu_L$ would be indistinguishable from
$\nu_R$.But in an orientable universe they would be distinct
particles.}
\medskip

\noindent {\bf 4.Composite $\nu_L\!{-}\!\nu_R$ system with interaction}
\medskip
 
Consider now an interaction $V({\bf x,y})$ between $\nu_L$ 
and $\nu_R$ of the form 
$$
V({\bf x,y}) = F(|{\bf x}|)H(|{\bf x}|-|{\bf y}|) + F(|{\bf y}|)H(|{\bf y}|-|{\bf x}|) \eqno(15)
$$
where $F$ is a function to be specified and $H$ is the
Heaviside function.Thus
$$
\left. \eqalign{
V({\bf x,y}) = & F(|{\bf x}|) \quad if \quad |{\bf x}| > |{\bf y}| \cr
& F(|{\bf y}|) \quad if \quad |{\bf x}| < |{\bf y}|}\right\}
\eqno(16)
$$ 
Note that $V({\bf x,y}) = V({\bf y,x})$. The combined 
Hamiltonian $H = H_o + V({\bf x,y})$ with total
energy say $E_{Tot}$ satisfies an equation of the form:
$$
[H_L + H_R + V({\bf x,y})] \Psi = E_{Tot} \Psi \eqno(17)
$$
We can now follow an approximation (i.e. perturbation)
procedure similar to that of the He-atom [6].(Although it
would be perhaps more appropriate to use the Lippmann-Schwinger
equation for the continuous case). Thus as a first approximation
the total energy is given by
$$
E_{Tot} = E_L + E_R + ( J \pm  K) \eqno(18)
$$
where
$$
J=<\Psi_1|V({\bf x,y})|\Psi_1>=<\varphi_{E_L}({\bf x}) \otimes \chi_{E_R}({\bf y})|V({\bf x,y})|\varphi_{E_L}({\bf x})
 \otimes \chi_{E_R}({\bf y})> \eqno(19)
$$
$$
K=<\Psi_2|V({\bf x,y})|\Psi_1>=<\varphi_{E_R}({\bf x}) \otimes\chi_{E_L}({\bf y})|V({\bf x,y})|\varphi_{E_L}({\bf x})
\otimes \chi_{E_R}({\bf y})> \eqno(20)
$$ 

\noindent We shall now suppose that $V$ is such that $J$ is non-positive, i.e.
$J(E_L,E_R)=-\Gamma (E_L,E_R)$, where $\Gamma (E_L,E_R)
\geq 0$ for $E_L,E_R \geq 0$ (see example later), so that
$$
E_{Tot}=E_L + E_R - (\Gamma \pm |K|) \eqno(21)
$$

\noindent The total energy of such an interactive system
thus has two branches. Note that both $\Gamma$ and $|K|$ are
functions of $E_L$ and $E_R$. If we start with a $\nu_L$ 
of energy $E_L > 0$ and a $\nu_R$ of energy $E_R > 0$ and 
switch on the interaction, the total energy $E_{Tot}$ would
be positive only if $E_L + E_R > \Gamma \pm |K|$. The value
of $E_L + E_R$ where $E_{Tot} = 0$ and changes sign, can be
interpreted as the rest-mass of a free (stable) particle. 

\noindent {\bf 5.A reduced 1-dimensional model}
\medskip

In order to
evaluate $J$ and $K$ for any specific model we need first 
to evaluate the surface integrals (10) and (12) over the
energy shells $S^2_{E_L}$ and $S^2_{E_R}$.
\medskip

\noindent Setting $p_3=\pm\sqrt{E^2_L - (p^2_1 + p^2_2)}$
and introducing the polar coordinates $(r,\theta)$ in the 
$p_1\!{-}\!p_2$ plane, we get $\varphi_{E_L}({\bf x})=\pmatrix{
\varphi^1_{E_L}({\bf x}) & \cr \varphi^2_{E_L}({\bf x})}$,
where
$$
\varphi^1_{E_L}({\bf x})=
\int _{0}^{E_L}\left (\int^{2\pi}_{0}e^{i(r a(\theta ) - \theta )} d \theta \right )
\!2\,{\frac {\cos(x_3\sqrt {{E_L}^{2}-{r}^{2}})E_L{r}
^{2}}{\sqrt {{E_L}^{2}-{r}^{2}}}}{dr} \eqno(22a)
$$
$$
\left .\eqalign{
\varphi^2_{E_L}({\bf x}) = & \int _{0}^{E_L}\left (\int^{2\pi}_{0}e^{ir a(\theta )} d \theta \right ) \times \cr 
 & \!2\,{\frac {
\left (E_L\cos(x_3\sqrt {{E_L}^{2}-{r}^{2}})-i
\sqrt {{E_L}^{2}-{r}^{2}}\sin(x_3\sqrt {{E_L}^{2}-{r}^{2}})\right )E_Lr}{\sqrt {{E_L}^{2}-{r}^{2}}}}{dr}
}\right\} \eqno(22b)
$$
Here $a(\theta )= x_1 \cos (\theta ) + x_2 \sin (\theta )$.
\medskip

\noindent Unfortunately, the above integrals cannot be
evaluated in closed form. Consider therefore a reduced
one-dimensional model with $x_1=x_2=0$. Then $a(\theta) = 0$,
so that (setting $x_3=x$ )
$$
\varphi_{E_L}(x)=\pmatrix{0 & \cr
\cr
4\,{\frac {\left (E_L\sin(E_L x)x-i\sin(E_Lx)+ixE_L\cos(E_L x)\right )E_L
\pi }{{x}^{2}}}} \eqno(23)
$$
$\varphi_{E_L}(x)$ is square integrable , that is
$$
<\varphi_{E_L}(x)|\varphi_{E_L}(x)>=
\int^\infty_{-\infty}\varphi_{E_L}(x)\dagger\varphi_{E_L}(x)dx=(\textstyle{64\over3})\pi^3 E^5_L \eqno(24)
$$
\medskip
\medskip
\noindent The normalizing factor for $\varphi_{E_L}(x)$ is thus
$N_{E_L}= \left ( (\textstyle{64\over3})\pi^3 E^5_L \right )^{-1/2}$.
From now on we shall suppose that $\varphi_{E_L}(x)$ is 
normalised.
\medskip

\noindent Similarly, setting $y_1=y_2=0,y_3=y$ we get from (12)
$$
\chi_{E_R}(y)=\pmatrix{
{\frac {4\,i\left (\sin(E_Ry)-E_Ry\cos(E_Ry)\right )\pi \,E_R}{{y}
^{2}}} & \cr
\cr
4\,{\frac {\sin(E_Ry)\pi \,{E_R}^{2}}{y}}
} \eqno(25)
$$
and
$$
<\chi_{E_R}(y)|\chi_{E_R}(y)>=\int^\infty_{-\infty}
\chi_{E_R}(y)\dagger\chi_{E_R}(y)dy=(\textstyle{64\over3})\pi^3 E^5_R \eqno(26)
$$
so that the normalising factor for $\chi_{E_R}(y)$ is also
$N_{E_R}= \left ( (\textstyle{64\over3})\pi^3 E^5_R \right )^{-1/2}$.
Again suppose that from now on $\chi_{E_R}(y)$ is
normalised.
\medskip

\noindent According to (19)
$$
J=\int\int \varphi_{E_L}(x)\dagger\varphi_{E_L}(x)V(x,y)
\chi_{E_R}(y)\dagger\chi_{E_R}(y)dxdy \eqno(27)
$$
Using a mean-value theorem for integrals we can write
$$
J=\varphi_{E_L}(\xi)\dagger\varphi_{E_L}(\xi)\chi_{E_R}(\eta)\dagger\chi_{E_R}(\eta)
\int\int V(x,y)dxdy \eqno(28)
$$
where $-\infty<\xi,\eta<\infty$.
\medskip

\noindent Now {\it assume} that $V$ is given by (16),where
the function $F$ is, for $x>0$ :
$$
F(x)=e^{-\mu x}ln(x) \eqno(29)
$$
\medskip
\noindent
Here $\mu > 0$ is some fundamental interaction constant.

 \hskip .5in{\epsfbox{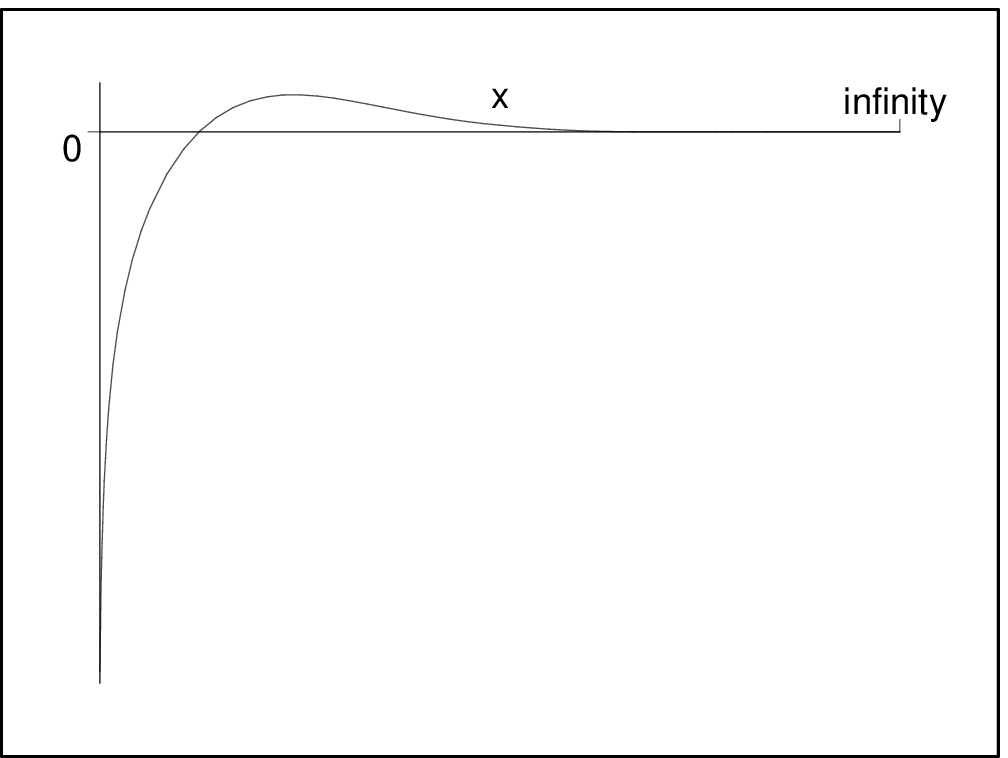}}
\medskip
\centerline{ Fig.1 }
\medskip
\noindent Fig.1 shows $ F(x)$ as a function of $ x $.
\medskip
\noindent
Then
$$
\left . \eqalign{
\int\int V(x,y)dxdy= & 4\int^\infty_0dy\left [ \int^y_0F(y)dx +\int^\infty_y F(x)dx \right] \cr
= & -8\left ( ln(\mu) + \gamma -1 \right )/\mu^2 \equiv -
\Gamma_o }\right \}
\eqno(30)
$$
Here $\gamma = 0.5772156649..$ (Euler constant). Hence
$\Gamma_o > 0$ if $\mu > e^{1-\gamma} = 1.526205112..$. We
shall suppose that this condition is satisfied, so that
$J=-\Gamma \leq 0$, where
$$
\Gamma(E_L,E_R)=\varphi_{E_L}(\xi)\dagger\varphi_{E_L}(\xi)\chi_{E_R}(\eta)\dagger\chi_{E_R}(\eta)\Gamma_o \eqno(31)
$$

\noindent Similarly
$$
K(E_L,E_R) = -\varphi_{E_R}(\vartheta)\dagger\varphi_{E_L}(\vartheta)\chi_{E_L}(\zeta)\dagger\chi_{E_R}(\zeta)\Gamma_o \eqno(32)
$$
where $-\infty<\vartheta,\zeta<\infty$.
\medskip

\noindent In order to obtain a qualitative idea of how
the two branches of $E_{Tot}$ behave as a function of
$E_L$ and $E_R$, consider the case where $E_L=E_R=E$.
Then $\Gamma = |K|$ and for the branch $E_{Tot} = 2E - (\Gamma+ |K|)$ we have, from (23),(25) and (31)
$$
\left . \eqalign{
E_{Tot}= 2E - & 3/4\,{\frac {{E}^{2}{\xi}^{2}+1-\left (\cos(E\xi)\right )^{2}-2\,\sin(E\xi)\xi E\cos(E\xi)}{\pi \,{E}^{3}{\xi}^{4}}}
\times \cr
& 3/4\,{\frac {{E}^{2}{\eta}^{2}+1-\left (\cos(E\eta)\right )^{2}-2
\,\sin(E\eta)\eta E\cos(E\eta)}{\pi \,{E}^{3}{\eta}^{4}}}
\Gamma_o}\right \} \eqno(33)
$$
\vfil\eject
\hskip .5in{\epsfbox{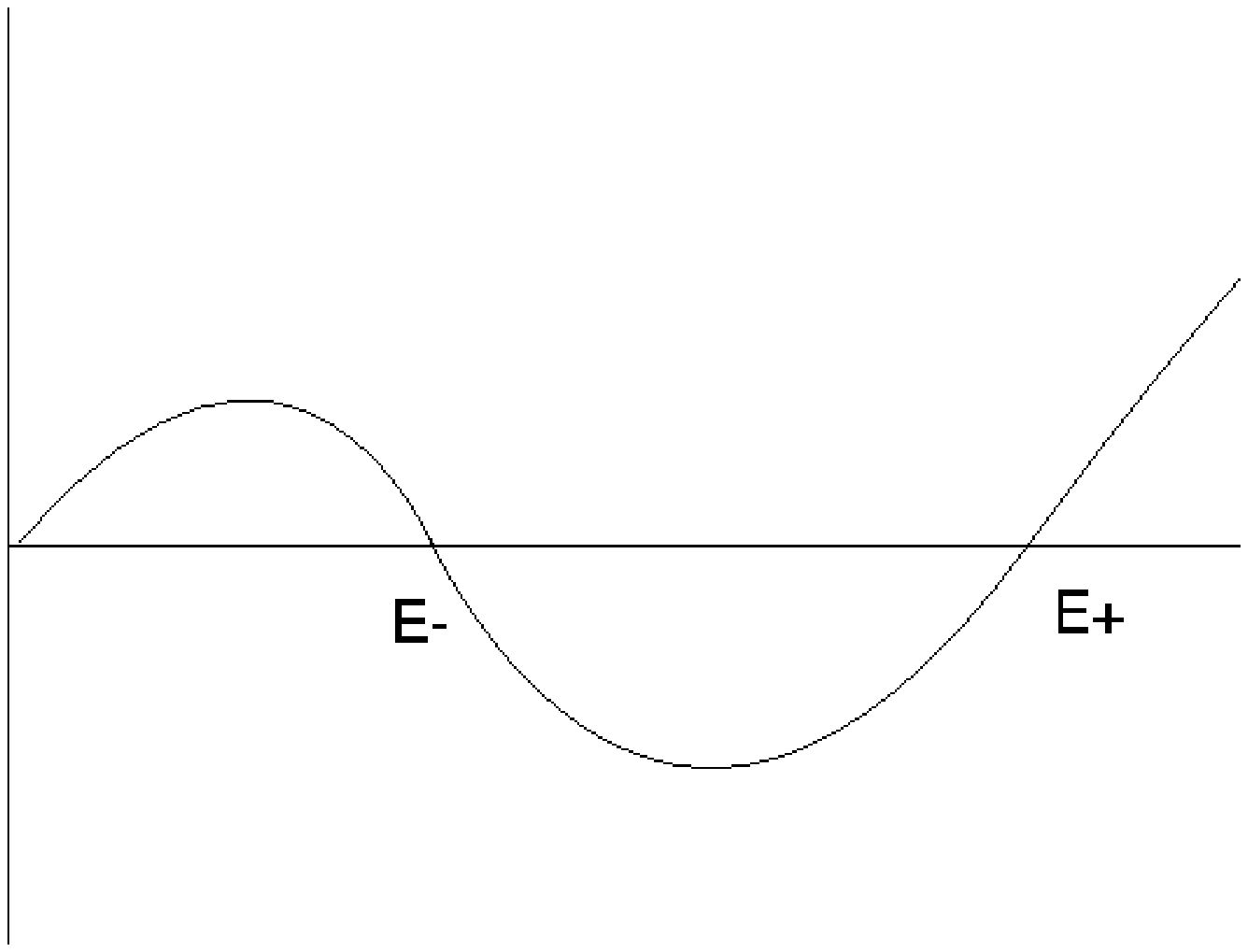}}
\medskip
\medskip 
\centerline {Fig.2 $E_{Tot}$ as a function of $E$}
\medskip
\noindent Fig.2 is a plot of $E_{Tot}$ as a function of
$E$ for some specific values of $\xi,\eta,\Gamma_o$. It shows that, in general,
for values $0<E<E-$, $E_{Tot}$ is positive; it
becomes zero at $E=E-$; it is then negative when $E- < E < E+$. It is positive again when $E > E+$. 
\medskip
\noindent How can one interpret this somewhat strange behaviour of $E_{Tot}$ as a function of $E$?
\medskip
\noindent A possible interpretation is as follows.
\medskip
\noindent Now imagine that we start with a free  $\nu_L$ and a free
 $\nu_R$ both of energy $E$.
The free system has a mirror symmetry. If we now switch on the
above mentioned interaction, the symmetry is broken because
of the nature of the Heaviside function in the interaction.
Then depending on if $0<E<E-$ the composite system
is a 4-component Dirac neutrino of energy $E_{Tol} > 0$. If 
however $E- < E < E+ $ the composite system would contribute to vacuum energy $E_{Tot} < 0 $. If $ E > E+ $ we have a positive energy
defect of $ 2 E+ $. This can be interpreted as the formation of a 
particle of rest mass equal to $ 2 E+ $ and $ E_{Tot} $ now
representing the pure kinetic energy of the formed particle. 

\medskip
\noindent The rest mass $ 2 E+ $ would depend on the value
of the fundamental constant $\mu$.
\medskip
\noindent A similar analysis of the other branch (with $ E_L
\neq E_R $ ) would lead to another particle with rest mass
less than the above case.
\medskip

\noindent {\bf 6.Conclusion}
\medskip
\noindent We have thus shown that a simple quantum mechanical
analysis of a composite $\nu_L\!{-}\!\nu_R$ model with a specific
symmetry-breaking 
interaction suggests a possible formation of particles of non-
zero rest  mass from 2-component Weyl neutrinos of sufficiently high
energy. In this interactive model, in a model universe filled with
Weyl neutrinos $\nu_L$ and $\nu_R$ with energy spectrum 
$ 0 < E_L,E_R < \infty $, we can therefore expect { \it at least }
three things:
4-component Dirac neutrinos, a vacuum filled with negative energy and
two kinds of stable particles of non-zero rest mass.
\medskip
\noindent The analysis presented here is in the framework of
quantum mechanics of particles.
The next step would be to consider such a
model in the framework of quantum field theory.

\bigskip
\bigskip 
\centerline {\bf References}
\vglue .25truein

\baselineskip=12pt

\item{[1]\ } R. D. McKewon, P. Vogel: {\it Phys. Reports},
{\bf 394}, 315-356 (2004)
\medskip

\item{[2]\ } W. Heisenberg: {\it Rev. Mod. Phys},{\bf 29}, 269
(1957). 
\medskip

\item{[3]\ } W. Heisenberg {\it et al}: {\it Zeit. f. Naturforschung}, {\bf 14a}, 441 (1959). 
\medskip

\item{[4]\ } D. K. Sen: {\it Nouvo Cimento}, {\bf 31}, 660 
(1964).
\medskip

\item{[5]\ } Dipak  K. Sen: {\it J. Math. Phys.}, {\bf 48},
022304 1-8 (2007).
\medskip

\item{[6]\ } F. Schwabl: {\it Quantum Mechanics}, Springer,
 Berlin, Heidelberg (2002).

 \end